# A Comprehensive Study and Performance Comparison of M-ary Modulation Schemes for an Efficient Wireless Mobile Communication System

Md. Emdadul Haque[1], Md. Golam Rashed[2], M. Hasnat Kabir[3]

[1,3]Department of Information Communication Engineering,
University of Rajshahi, Rajshahi-6205, Bangladesh.
`hasnatkabir@yahoo.com`
[2]Department of Electronics and Telecommunication Engineering (ETE),
Prime University, Dhaka-1209, Bangladesh
`golamrashed.ru@gmail.com`

### ABSTRACT

*Wireless communications has become one of the fastest growing areas in our modern life and creates enormous impact on nearly every feature of our daily life. In this paper, the performance of M-ary modulations schemes (MPSK, MQAM, MFSK) based wireless communication system on audio signal transmission over Additive Gaussian Noise (AWGN) channel are analyzed in terms of bit error probability as a function of SNR. Based on the results obtained in the present study, MPSK and MQAM are showing better performance for lower modulation order whereas these are inferior with higher M. The BER value is smaller in MFSK for higher M, but it is worse due to the distortion in the reproduce signal at the receiver end. The lossless reproduction of recorded voice signal can be achieved at the receiver end with a lower modulation order.*

## 1. INTRODUCTION

Wireless services have been growing firstly in each year [1]. Present third generation telecommunication system provides a more flexible data rate, a higher capacity and a tightly integrated service. Transmit a unit information is the main objective of a communication system [2]. Modulation and demodulation are the basic features for transmitting and receiving information in wireless system. Thus, modulator is a fundamental component in wireless equipment. Modern mobile and wireless communication systems adopt digital modulation scheme as an essential module for signal processing instead of previously used analog modulation. Digital modulation is very much advantageous in noise immunity and robustness to channel impairments. In cellular wireless communication, the reuse of the same allocated frequency spectrum in different geographical areas fascinates the development of large scale cellular mobile networks. Usually transmission of audio signal has to pay enormous cost of bandwidth and need to have lossless reproduction at the receiving end. Therefore, it is meaningful to investigate the performance of a wireless mobile communication system with the deployment of M-ary ( MPSK, MQAM, MFSK ) modulation schemes for audio signal. These improved and most promising digital modulation schemes are widely used due to their increased spectral efficiency [3-5]. The performance of a communication system depends on several factors, determining the bit error rate of the technique is one of them [6, 7]. The objective of this paper is to review the key performance of the main M-ary modulation schemes and compares them to each other according to their performance such as bit error rate in the presence of Additive White Gaussian Noise (AWGN).

## 2. COMMUNICATION SYSTEM MODEL

Figure 1 shows the block diagram of presented communication system model utilizing M-ary (MPSK, MQAM, MFSK) modulation scheme. This is a point-to point communication system model [8]. The figure indicates the step by step operation of each functional block. At the transmitter side, the input binary data stream obtained from the recorded voice signal. The retrieved binary data stream is passed through the other required steps (source of encoder, channel coder) as the signal has to transmit. Our main observation is concentrated to the next block (M-array modulator) of the considered communication system model where we were planned to vary the modulation order of different modulation schemes to analysis their performances. The performance and characteristics of a system is significantly depended on the choice of digital modulation scheme. There is no basic rule for choosing a scheme. However, one scheme is better than other depending of the channel, required levels of performance and target hardware trade-offs [9]. Nevertheless, it also must be considered the required data rate, latency, link budget as well as available bandwidth. At the receiver side, the received signal is demodulated, and passed through the other subsequent blocks in order to reproduce the transmitted speech signal.

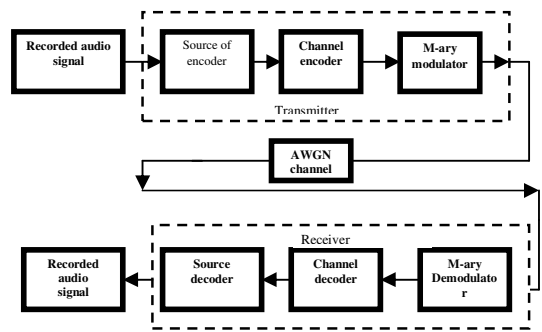

Figure 1 Block diagram of digital communication system based on M-ary (MPSK, MQAM, MFSK) modulation scheme.

## 3. ANALYSIS OF THE M-ARY SYSTEMS

The performance of M-ary phase shift keying (PSK), M-ary quadrature amplitude modulation (QAM) and M-ary frequency shift keying (FSK) for a considered wireless communication system are evaluated in terms of bit error probability. The expressions of error probability are derived using bounds and approximations to analyze and compare the performance.

### 3.1. Error Probability of MPSK

The bit error probability $P_{MPSK}$ for M-ary PSK modulation scheme is given by [10, 11]

$$P_{MPSK} = 1 - \frac{1}{2\pi} \int_{-\frac{\pi}{M}}^{\frac{\pi}{M}} e^{-\lambda} \left[1 + \sqrt{4\pi\lambda} \cos\theta \times e^{\lambda \cos^2\theta} \left(1 - Q\left(\sqrt{2\lambda} \cos\theta\right)\right)\right] d\theta, \ldots \ldots \quad (1)$$

Where $\lambda = \dfrac{E_s}{N_o}$ is the signal to noise ratio per bit (SNR/bit), modulation order $M = 2^k$ as each symbol conveys k bits and the error function is Q.

But we can not say the Eq.(1) is a closed form. Therefore, we have to calculate it numerically, which is not so easy. To do this, we have to take some approximations to calculate the error probability. If we apply union boundary conditions and variants then the Eq.(1) can be written as

$$P_{MPSK} = 2Q\left(\sqrt{2\lambda \sin^2 \tfrac{\pi}{M}}\right) - \dfrac{1}{\pi}\int_{\tfrac{\pi}{2}-\tfrac{\pi}{M}}^{\tfrac{\pi}{2}} \exp\left(-\lambda \dfrac{\sin^2 \tfrac{\pi}{M}}{\cos^2 \theta}\right) d\theta \quad\ldots\ldots\ldots\ldots \quad (2)$$

which gives

$$P_{MPSK} \cong 2Q\left(\sqrt{2\lambda \sin^2 \tfrac{\pi}{M}}\right). \text{ (1}^{st}\text{ approx)}\ldots\ldots\ldots\ldots\ldots\ldots \quad (3)$$

and for large value of M, Eq.(3) can be written as

$$P_{MPSK} \cong 2Q\left(\sqrt{2\lambda \tfrac{\pi^2}{M^2}}\right), \text{ (2}^{nd}\text{ approx)}\ldots\ldots\ldots\ldots\ldots\ldots\ldots\ldots \quad (4)$$

The closed forms of Eq.(1) are equations (3) and (4) for first and second approximation, respectively. These equations can be applied to calculate the error probability in terms of SNR per bit. It is necessary to increase the transmitted power in order to maintain the same performance level for higher M.

## 3.2. Error probability of MQAM

The bit error probability $P_{MQAM}$ for M-ary QAM modulation scheme is given by [10, 11]

$$P_{MQAM} = 1 - \left\{1 - 2\left[1 - \dfrac{1}{\sqrt{M}}\right]Q\left[\sqrt{\dfrac{3\lambda}{(M-1)}}\right]\right\}^2 \quad\ldots\ldots\ldots\ldots\ldots\ldots\ldots \quad (5)$$

To simplify this equation we can employ some further approximations in order to calculate the error probabilities, which are listed below.

$$P_{MQAM} \cong 4\left(1 - \dfrac{1}{\sqrt{M}}\right)Q\left(\sqrt{\dfrac{3\lambda}{(M-1)}}\right), \text{ (1}^{st}\text{ approx)}\ldots\ldots\ldots\ldots\ldots \quad (6)$$

$$P_{MQAM} \cong 2\left(\dfrac{M}{M-1}\right)Q\left[\sqrt{\dfrac{6\lambda}{(M^2-1)}}\right], \text{ (2}^{nd}\text{ approx)}\ldots\ldots\ldots\ldots\ldots. \quad (7)$$

and for large value of M

$$P_{MQAM} \cong 2Q\left[\sqrt{\dfrac{6\lambda}{M^2}}\right], \text{ (3}^{rd}\text{ approx)}\ldots\ldots\ldots\ldots\ldots\ldots\ldots\ldots\ldots. \quad (8)$$

## 3.3. Error probability of MFSK

The bit error probability $P_{MFSK}$ for M-ary FSK modulation scheme is given by [8, 12]

$$P_{MFSK} = 1 - \frac{1}{\sqrt{2\pi}} \int_{-\alpha}^{\alpha} \exp\left[\left\{-\frac{1}{2}(y-\sqrt{2\lambda})^2\right\}\right][1-Q(y)]^{M-1} dy, \ldots\ldots\ldots\ldots\ldots \quad (9)$$

But Eq.(1) is not a closed form. Therefore, we have to calculate it numerically. To do this, we have to take some approximations to calculate the error probability. If we apply union boundary conditions and variants then the Eq.(1) can be written as

$$P_{MFSK} \cong \frac{M \log_2 M}{2} Q[\sqrt{\lambda}], \text{(approx)}\ldots\ldots\ldots\ldots\ldots\ldots\ldots\ldots\ldots \quad (10)$$

MFSK shows in contrast to the other modulation schemes, the error probability decreases as M increases.

## 4. SIMULATION RESULTS AND COMPARISON

M-ary modulations schemes in wireless communication systems are analyzed using voice signal as a function of bit error rate (BER). This is a simulation work dome by Matlab. The work deals the recorded voice signal which is transmitted though AWGN channel where fading channels like Rayleigh fading and Ricean fading channel are not considered.

Figure 2 illustrates an image of recoded voice signal. The signal has been downloaded from the website. This signal is used as transmitted signal and reproduces it at the receiver end through out the whole work for different values of signal to noise ratio ranging from 0 to 12 dB.

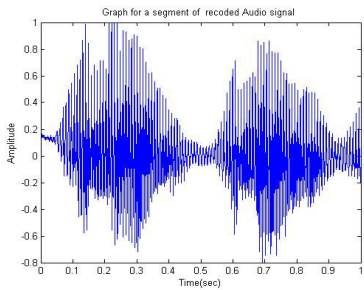

Figure 2: Original recorded voice signal

Figure 3 shows the comparison of MPSK, MFSK and MQAM as a bit error probability for modulation order M=2. The bit error rate is sharply decreased with respect to signal to noise ratio. From this figure, it is clear that, the performance of MPSK is better than that of others. However, MPSK and MQAM show almost smaller result whereas MFSK shows higher bit error rate. More that one order of magnitude higher BER value can be found in MFSK at the SNR=6 dB than that of other modulation techniques. Reproduction of original voice signal with modulation order M=2 is shown in Figure 4. As we mention earlier that MPSK and MQAM show similar value which is also found in the reproduced signals ( 4.a and 4.b). On the other hand, figure 4.c shows distortion in the reproduced signal of MFSK.

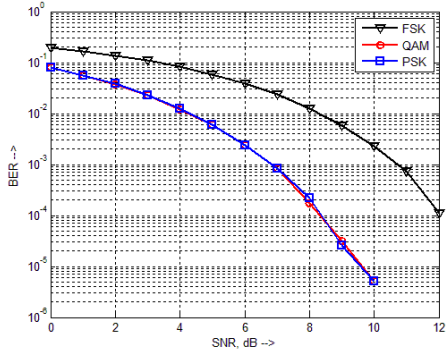
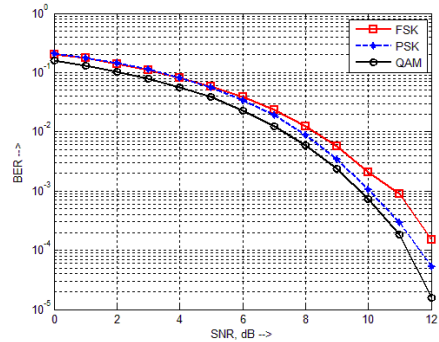

Figure 3: System performance under different modulation schemes (MPSK, MQAM, and MFSK) for M=2.

Figure 5: System performance under different modulation schemes (MPSK, MQAM, and MFSK) for M=4.

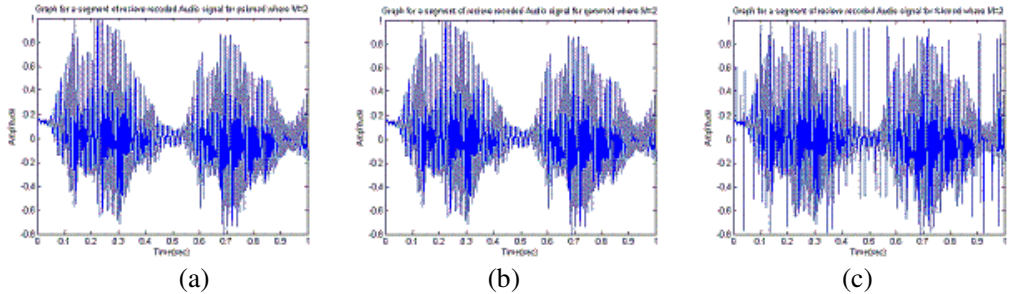

(a)      (b)      (c)

Figure 4: Reproduction of transmitted voice at receiving end for the value of EbNo=6dB after transmitting through AWGN channel [a) MPSK, b) MQAM, c) MFSK for M=2].

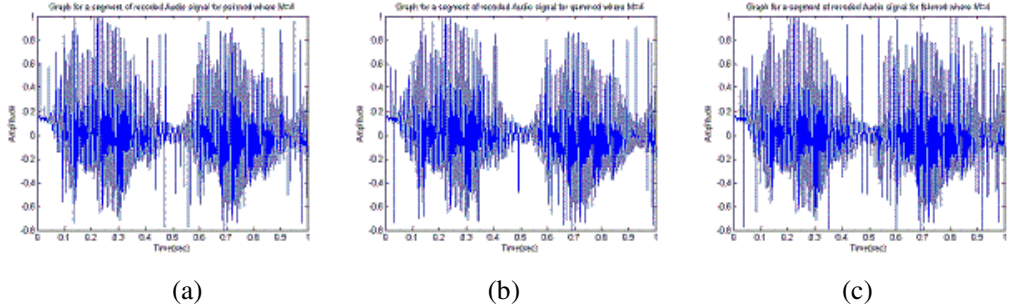

(a)      (b)      (c)

Figure 6: Reproduction of transmitted voice at receiving end for the value of EbNo=6dB after transmitting through AWGN channel [a) MPSK, b) MQAM, c) MFSK for M=4].

Figure 5 represents that the performance of MQAM is better in contrast to MPSK and MFSK for the modulation order M=4. Reproduction of original signal is shown in figure 6. From figure 4 and 6, it is found that MFSK shows worst value. Therefore, it can be pointed out that the MFSK technique is not suitable for lower modulation order for transmitting audio signal. On the other hand, MFSK achieves better performance in contrast to other modulation schemes according to bit error rate with increasing the value of modulation order that is shown in figure 7. Similar result has been found elsewhere [4]. The reproduction original signal has huge distortion shown in figure 8 suggest that the higher modulation order is not suitable here even though the BER is smaller.

International Journal of Computer Science, Engineering and Applications (IJCSEA) Vol.1, No.3, June 2011

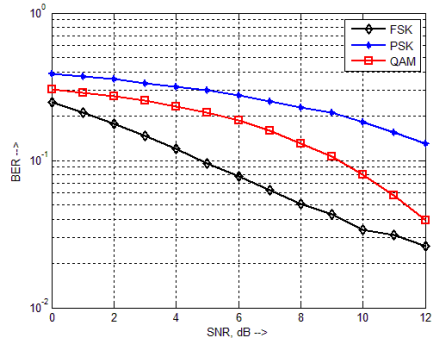

Figure 7: System performance under different modulation schemes (MPSK, MQAM, and MFSK) for M=16.

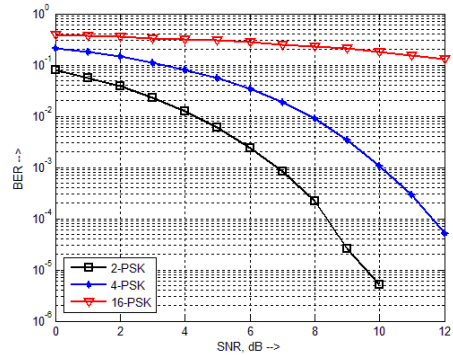

Figure 9: The performance of MPSK modulation schemes under different modulation orders.

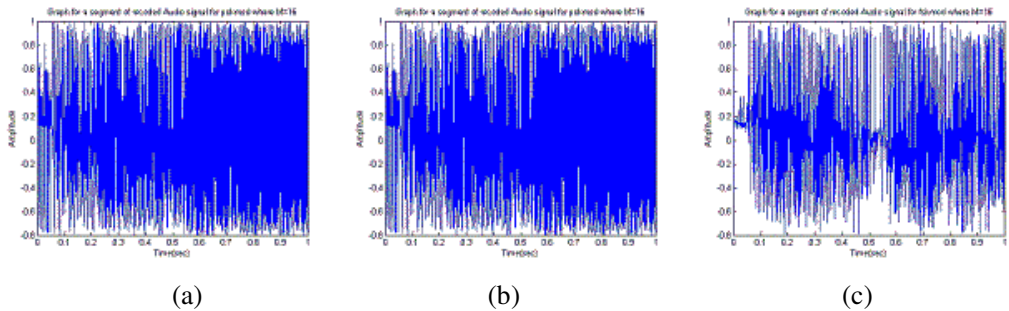

(a)      (b)      (c)

Figure 8: Reproduction of transmitted voice signal at receiving end for the value of EbNo=6dB after transmitting through AWGN channel [a) MPSK, b)MQAM, c)MFSK for M=16].

However, in compression between figure 3 and 5, it is clearly observed that for a fixed SNR value (SNR=6 dB), MPSK shows better BER performance than that of MQAM. Because, approximately 9 and 14 times lower BER value can be found in MQAM and MPSK scheme, respectively for modulation order M=2 than that of M=4. Therefore, we can say that MPSK outperform when the modulation order M=2. We also found that the BER value is directly proportional to the modulation order shown in figure 9.

From Figure 3 , Figure 5 and Figure 7 , it is also observed that, in order to maintain a given performance level, it is possible to increase the transmitted power as in the MPSK and MQAM or the transmitted bandwidth as in the MFSK for recorded audio signal through a wireless communication system.  Because the transmission bandwidth is fixed in modulation techniques MPSK and MQAM, but it increases exponentially by $2^{2k}$ as M increases. The transmitted power, on the other hand, in MFSK modulation technique is basically independent of k, but the transmitted bandwidth increases exponentially by $2^k$ as M increases.

## 5. CONCLUSION

The performances of M-ary modulation schemes for wireless and mobile communication system under AWGN channel are evaluated using bit error probability in order to transmit the recorded voice signals. Based on the results obtained in the present study, it can be seen that the MPSK and MQAK are showing better performance for lower modulation order whereas these International Journal of Computer Science, Engineering and Applications (IJCSEA) Vol.1, No.3, June 2011are inferior with higher M. Therefore, it can be concluded that the lossless reproduction of recorded voice signal can be achieved at the receiver end with a lower modulation order.

**REFERENCES**

[1] T. S. Rappaport, "Wireless Communications Principles & Practice", IEEE Press, New York, Prentice Hall, pp. 169-177, 1996.

[2] Md. Golam Rashed, M. Hasnat Kabir, Sk. Enayet Ullah and Rubaiyat Yasmin, "Performance evaluation of CRC Interleaved QPSK based Wireless Communication System for Color Image Transmission." Journal of Bangladesh Electronics Society (BES), Dhaka, Bangladesh. December/2009.

[3] Hongzhi Zhao, Yi Gong, Yong Liang Guan, and Youxi Tang, "Performance Analysis ofM-PSK/M-QAM Modulated Orthogonal Space–Time Block Codes in Keyhole Channels" IEEE Transactions on Vehicular Technology, Vol. 58, No. 2, February 2009.

[4] M.A. Razzak, F. Ndiaye, O. Khayam, A. Jabbar, "On the Performance of M-ary Modulation Schemes for Efficient Communication Systems "10th international conference on Computer and information technology, 2007. iccit 2007.

[5] M. Riaz Ahmed, Md.Rumen Ahmed , Md.Ruhul Amin Robin, Md. Asaduzzaman, Md. Mahbub Hossain, Md. Abdul Awal, "Performance Analysis of Different M-Ary Modulation Techniques in Fading Channels using Different Diversity", Journal of Theoretical and Applied Information Technology, Vol. 15. No.1, pp 23-28, May 2010.

[6] T. L. Staley, Richard C. North, Jianxia Luo, Walter H. Ku, and James R. Zeidler," Performance Evaluation for Multichannel Reception of Coherent MPSK over Frequency Selective Fading Channels", IEEE Transactions on Vehicular Technology, Vol. 50, No. 4, July 2001.

[7] Thomas L. Staley, Jianxia Luo,Walter H. Ku, James R. Zeidler, "Error Probability Performance Prediction for Multichannel Reception of Linearly Modulated Coherent Systems on Fading Channels", IEEE Transactions on Communications, Vol. 50, No. 9, September 2002.

[8] John G. Proakis. "Digital Communications ," 3$^{rd}$ edition,. Singapore, McGraw-Hill Book Co.1995

[9] G. Smithson, "Introduction to digital modulation schemes", IEE Digest / Volume 1998 / Issue 240, doi:10.1049/ic:19980230

[10] C. M. Chie, "Bound and approximations for rapid evaluation of coherent MPSK error probabilities", IEEE Trans.Com., Vol. No. 3, pp. 271–273, 1985.

[11] Jianhau Lu, K. B. Letaif, Justin C-I Chuang and Ming L. Liou, "M-PSK and M-QAM BER Computation Using Signal- Space Concepts", IEEE Trans. Com., Vol. 47, No. 2, 1999.

[12] J. G. Proakis and M. Salehi, "Communication Systems Engineering", 2nd Ed., Pearson Education, 2002.
45